G Sun et al.

# Neutral pressure measurement in TCV tokamak using ASDEX-type pressure gauges

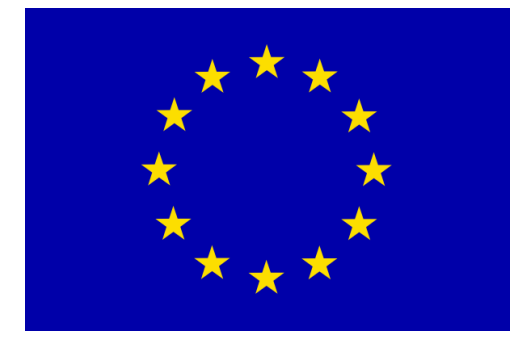

This work has been carried out within the framework of the EUROfusion Consortium and has received funding from the Euratom research and training programme 2014-2018 and 2019-2020 under grant agreement No 633053. The views and opinions expressed herein do not necessarily reflect those of the European Commission.

# Neutral pressure measurement in TCV tokamak using ASDEX-type pressure gauges

G. Sun[1], H. Reimerdes, H. Elaian, M. Baquero-Ruiz, B. Brown, M. Gospodarczyk, M. Noël, E. Tonello, and the TCV team[2]

*Ecole Polytechnique Fédérale de Lausanne (EPFL), Swiss Plasma Center (SPC), CH-1015 Lausanne, Switzerland*

Probing the neutral gas distribution at the edge of magnetic confinement fusion devices is critical for plasma exhaust studies. In the TCV tokamak, a set of ASDEX-type hot ionization pressure gauges (APGs) has been installed for fast, in-situ measurements of the neutral pressure distribution in the TCV chamber. The APGs have been calibrated against baratron pressure gauges (BPGs) for pressures ranging from less than 1 mPa to several hundred mPa. A correction to account for the residual pressure in the pumped torus is proposed to improve the measurement accuracy. APG measurements in a series of plasma discharges with varied density ramp rates are analyzed and compared with the BPG pressure measurements. APG measurements feature a significantly faster time response and extend the BPG measurement range to lower pressures. Systematically higher neutral pressures measured with APGs compared to BPGs connected to the same TCV port, are attributed to the BPG's slower time response and a nonuniform neutral distribution in gauge ports during the plasma discharge. The initial APG operations in TCV have been proven successful, which validates the APG as an adequate pressure measurement technique for the upcoming TCV divertor upgrades.

## I. INTRODUCTION

Neutral particles play an essential role in the operation of magnetic confinement fusion devices. Neutrals surrounding the plasma are directly linked to the plasma density and, thereby, the discharge regime and performance. Poloidally distributed gas valves and gas baffles can affect the poloidal distribution of neutrals. A higher divertor neutral pressure, $P_{n,div}$, generally leads to colder divertor plasmas and is favorable to access a detached divertor regime[1], an operation mode widely assumed to be necessary in future fusion reactors for tolerable power exhaust. Additionally, $P_{n,div}$ is related to the upstream separatrix density

[1] Author to whom any correspondence should be addressed. Author email: guang-yu.sun@epfl.ch
[2] See author list of B.P. Duval et al 2024 Nucl. Fusion 64 112023.

[2], which is limited by the global density limit [3]. In TCV, different shapes of neutral gas baffles have been installed to increase the divertor neutral confinement and achieve higher $P_{n,div}$ for the same core plasmas [4]. However, the neutral pressure surrounding the main plasma, $P_{n,main}$, should generally be low as neutrals can cool the confined plasma and lead to energetic charge exchange neutral that may damage the wall of fusion reactors. Precise and accurate measurements of the neutral pressure are, hence, crucial to validate the effectiveness of the gas baffles.

The neutral pressure used to be only measured with BPGs in TCV [5]. However, these BPGs suffer from a significant time delay and a slow time response, as they are connected to the TCV chamber via long extension tubes (of the order of two meters). Furthermore, the BPGs cannot resolve pressure levels below several mPa, expected in the main chamber of the baffled TCV divertor [6]. These limitations have motivated the installation ASDEX-type pressure gauges (APGs), which were specifically designed for in-situ, fast measurement of a wide range of neutral pressures in in the edge of magnetic confinement fusion devices [7].

APGs are pressure gauges designed for high magnetic field and noisy environments, which were developed by G. Haas and first deployed in ASDEX [7, 8]. APGs have since been installed in numerous tokamaks and stellarators, including ASDEX-Upgrade [9], DIII-D [10], EAST [11], KSTAR [12], ADITYA [13], LHD [14], W7-X [15] and are also foreseen for ITER [16]. The APG operation faces a range of technical challenges. The gauge filament, usually made of tungsten, is subjected to strong, cyclical J×B forces and thermal stresses, which may cause creep and fatigue of the filament [17]. In addition, APG measurement can suffer from small current jumps [18], and are affected by impurity neutrals [19] and time-varying magnetic field during the measurement [20]. All these concerns must be addressed to ensure long-term, stable and accurate APG measurements in fusion devices. The present work introduces the APG setup in TCV, highlights several encountered issues and proposes improvements that should be implemented, before employing APGs in the next TCV upgrade.

This article is structured as follows. Section II recaps the working principle of APG and introduces the APG setup in TCV. Section III provides the detailed procedure employed to calibrate TCV's APGs. Section IV compares the APG and BPG measurement in TCV plasma discharges with varied density ramp rates. Concluding remarks are given in section V.

## II. Diagnostic setup

This section briefly reviews the working principle of APGs (see [7] for more details) before presenting their setup in TCV, including the gauge head design, remote control, automation, and integration of the diagnostic into the TCV discharge cycle.

### A. Working principle of the ASDEX-type pressure gauge

The APG is a hot-cathode ionization pressure gauge [21] for environments with high magnetic fields. The gauge head consists of a filament, a control electrode (CE), an acceleration grid (AG), and an ion collector (IC), mounted on an electrically insulated base plate. The gauge head is oriented for its components to align with the magnetic field of the device, Figure 1(a). The filament is a metal wire usually made of tungsten and is biased positively. It is heated by an AC or DC current to emit a thermionic electron current. Filament shape and diameter are optimized for sufficient mechanical stability and a manageable heating current amplitude [17, 22]. The AG is a metallic plate with a slit that contains an acceleration grid, biased positively with respect to the filament to accelerate emitted electrons along magnetic field lines. The potential difference must be larger than the ionization energy of the neutral gas. The IC is located at the far side of the linear assembly and consists of a grounded metal plate, A CE, which is a metal plate featuring a narrower slit perpendicular to the magnetic field and parallel to the base plate, is placed between the filament and the AG. Its potential is varied between 0 and a potential between the filament and AG bias voltages, to chop the electron current at designated frequency.

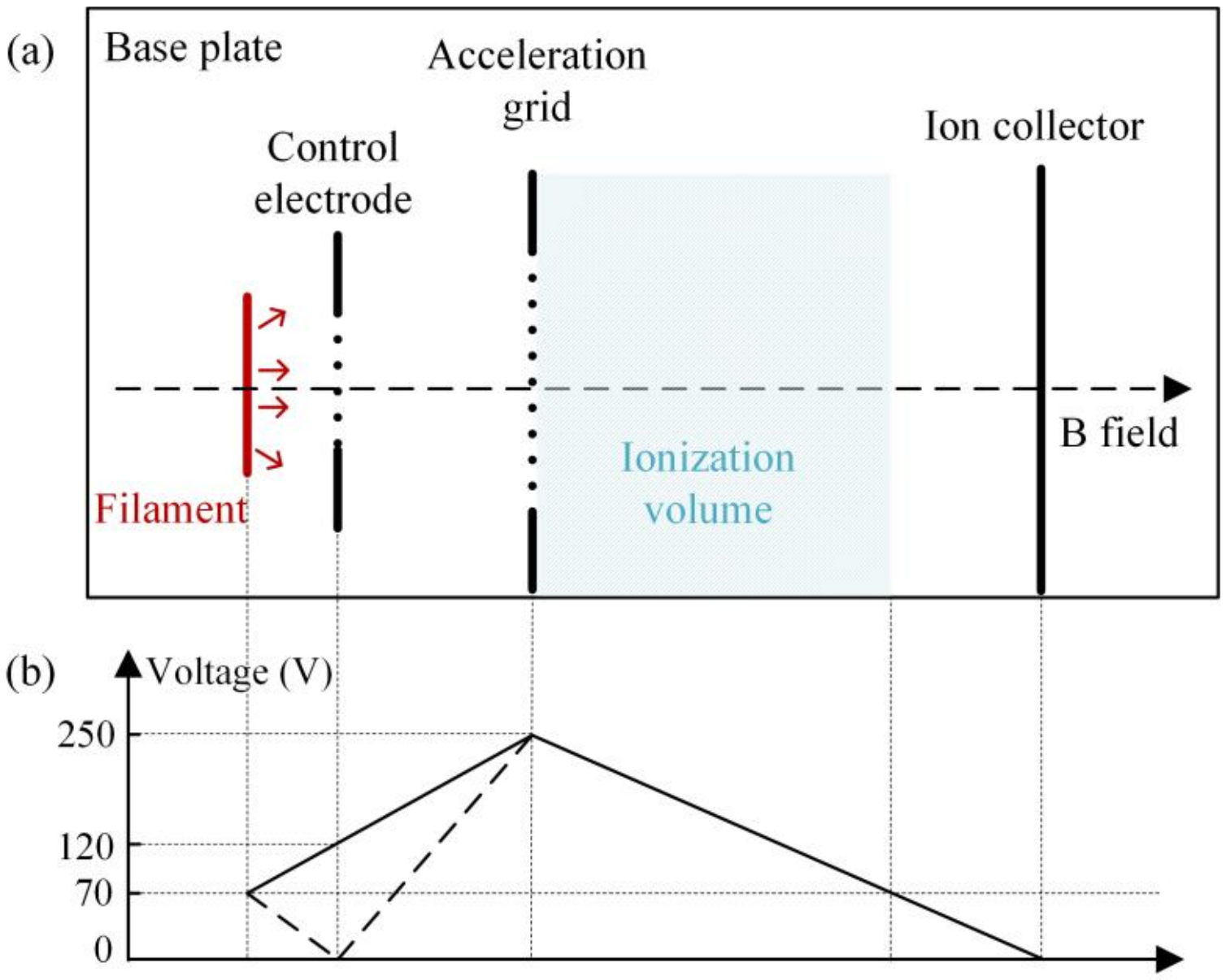

Figure 1. (a) Schematic top view of the APG head. Dotted lines in the control electrode and the acceleration grid are the slits. (b) The electrode biased voltages and the voltage distribution with dashed lines representing the voltage distribution during the control electrode chopping.

Electrons emitted from the heated filament are confined by the space potential between the filament and the IC, where they oscillate with sufficient kinetic energy to ionize neutrals, Figure 1(b). The ion current, $I^+$, resulting from electron-impact ionizations in the ionization volume, and electron current, $I^-$, are collected at the IC and AG, respectively. The neutral density in the ionization volume, Figure 1(a), can then be determined from the two measured currents, as $I^+$ increases approximately linearly with the neutral particle density in the ionization volume, $n_g$, and with $I^-$. This dependence is commonly expressed by the gauge's sensitivity, $d$, defined as:

$$d = \frac{I^+}{(I^- - I^+) n_g}. \quad (1)$$

The denominator of Equation (1) includes a correction for secondary electrons generated from ionization that contribute to $I^-$ but are typically not sufficiently energetic to ionize neutrals. A more physical expression for the APG sensitivity can be derived by calculating the ionization rate induced by oscillating electrons in the ionization volume at given potential distribution shown in Figure 1 (see Equations (1)-(10) in [7]):

$$d = \frac{2 \int_0^{e(U_{AG}-U_{Fil})} \sigma_{ion} \mathrm{d}\varepsilon \cdot x_{max} [1 + S(f-1)]}{e(U_{AG} - U_{Fil}) [\frac{1-\theta}{\theta} + S]}. \quad (2)$$

Here $\theta$ is the transparency of the AG, $S$ is the surviving probability (the probability for the electrons passing the AG to miss the filament when reflected by the electric field, due to e.g. collisions or drifts), $f$ is electron oscillation number, $x_{max}$ is the linear extent of the ionization volume in the direction parallel to the magnetic field, $U_{AG}$ and $U_{Fil}$ are voltages of AG and the filament, and $\sigma_{ion}$ is the electron-impact ionization cross section of the gas in the ionization volume. Since $\sigma_{ion}$ and, therefore $d$, depend on the gas species, the APG measurements in deuterium plasmas are affected by the presence of impurity neutrals, e.g. in discharges with impurity seeding. The gauge sensitivity is determined by in-situ gauge calibrations, section III.

With the gauge measuring the neutral density inside the gauge head, further considerations are needed to link this density to the neutral distribution in the vicinity of the gauge head. To facilitate the interpretation the APG head is shielded by a stainless-steel box with a small entrance hole to reduce the neutral conductance between the ionization volume and the region surrounding the gauge. This ensures that neutral particles inside the gauge undergo many wall collisions for atoms to recombine to molecules and these molecules to thermalize with the gauge shield before leaving the box. The neutral distribution outside the gauge is not necessarily at the same temperature as the gauge head, nor even a Maxwellian distribution. Since in stationary conditions, the influx of neutral particles through the entrance hole matches the outflux of thermalized molecules, the APG measurement for hydrogen isotopes can be interpreted as a measurement of the nuclei flux density towards the entrance hole:

$$\Gamma_n^{in} = \frac{1}{2} n_g \frac{\bar{v}_{th}}{4}, \quad (3)$$

where $\bar{v}_{th}$ is the mean velocity of the molecules in the gauge volume. This interpretation requires the knowledge of the gauge head temperature, which is assumed to be at room temperature. An increase of the gauge head temperature due to filament heating was verified to be negligible, by monitoring the temperature of an operating gauge for more than 10 s, typical for gauge operation on TCV, with a thermal imaging camera.

### B. APG setup in TCV

The APGs are set up to routinely operate during TCV discharges. The diagnostic is, therefore, automated and integrated in the general TCV discharge cycle, to minimize interventions by the diagnostic operator. The current APG setup in TCV is introduced below.

The APG diagnostic system in TCV consists of a National Instrument chassis containing an embedded PC and an FPGA card (NI-PXIe-7857R). The FPGA is programmed to execute the overall control of the APG measurement and controlled using LabVIEW, handling bias voltages, trigger and clock lines, signal conversion, data pre-processing and transfer, PID feedback control of the filament heating current, filament overheating protection, etc. It is connected to four gauge controllers (NGas modules[3]), which provide bias voltages to the filament, AG and CE, and measure the currents to AG and IC. The filament and AG are biased at 70 V and 250 V, respectively. and the CE voltage varies between 0 and 120 V with 5 kHz frequency. The filament is a thoriated tungsten wire with diameter of 0.6 mm, heated with a DC current of typically 20A. The heating current is feedback controlled for the AG to collect an electron current of typically 200 μA. Each NGas module is connected to a power supply (TDK-Lambda, GEN20-38) for filament heating, and to the gauge head where the measurement is performed[4], Figure 2(a).

The gauges are located in a port at the top (labeled 'top'), the high-field side floor (labeled 'div'), the low-field side floor (labeled 'bot'), and at the outer mid-plane (OMP) (labeled 'mic') of the TCV chamber, Figure 2(b). In typical lower-single null configurations the 'div' and 'bot' gauges measure the pressure in the private-flux region (PFR) and common-flux region (CFR) of the divertor, respectively. The gauges 'div', 'bot', and 'mic' are installed in ports that also connect to corresponding BPGs to facilitate the gauge commissioning. The electrical power to NGas modules and filament power supplies is controlled remotely. The filament power supplies are switched on for entire days with TCV operation, whereas the NGas modules are only switched on for each TCV discharge when human access to the tokamak is prohibited, due to the high bias voltages.

[3] The NGas modules were purchased from Arbeitsgruppe Weltraumphysik und -technologie (AWT).
[4] The gauge heads were purchased from IPT-Albrecht GmbH.

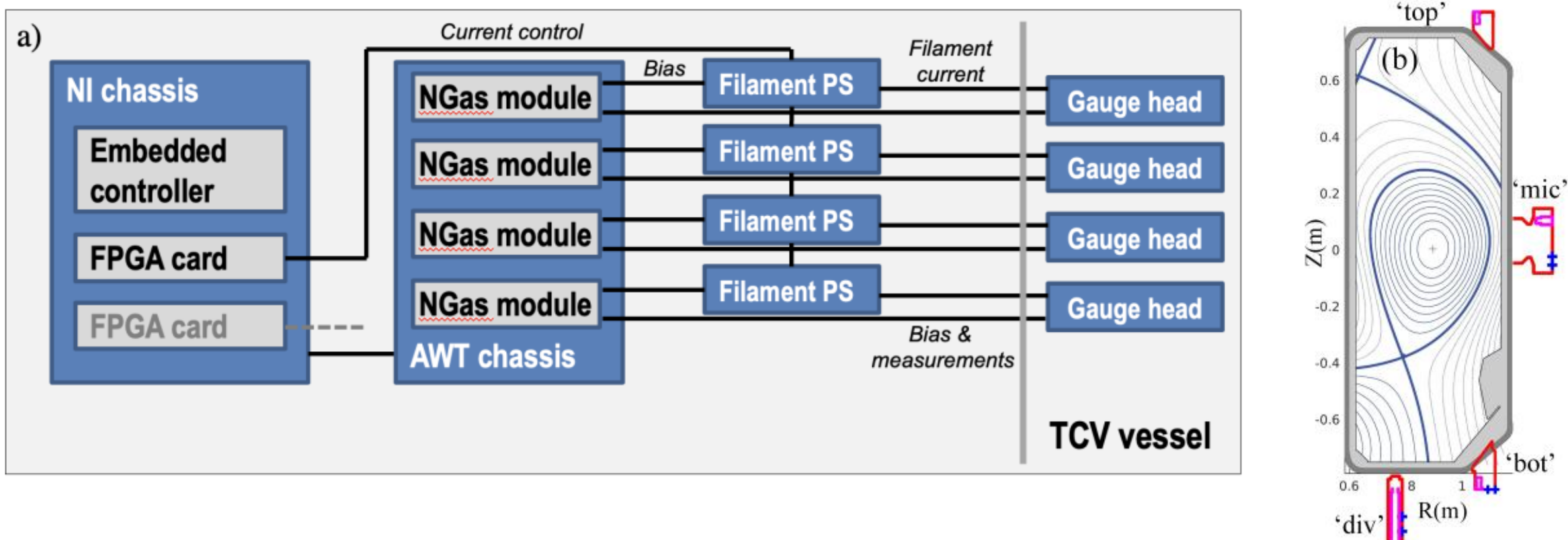

Figure 2. (a) Schematic of the APG measurement system on TCV. (b) Locations of the APG and BPG ports. The red contours are port structures containing the gauge heads, the magenta contours are the APG heads, and the blue points are opening for the tubes connected to the BPGs.

The leads of the gauge are covered by metallic circular tube to shield the signal cables against the potential influence of radiation on the current measurements, which is most obvious at the outer mid-plane gauge position, Figure 3.

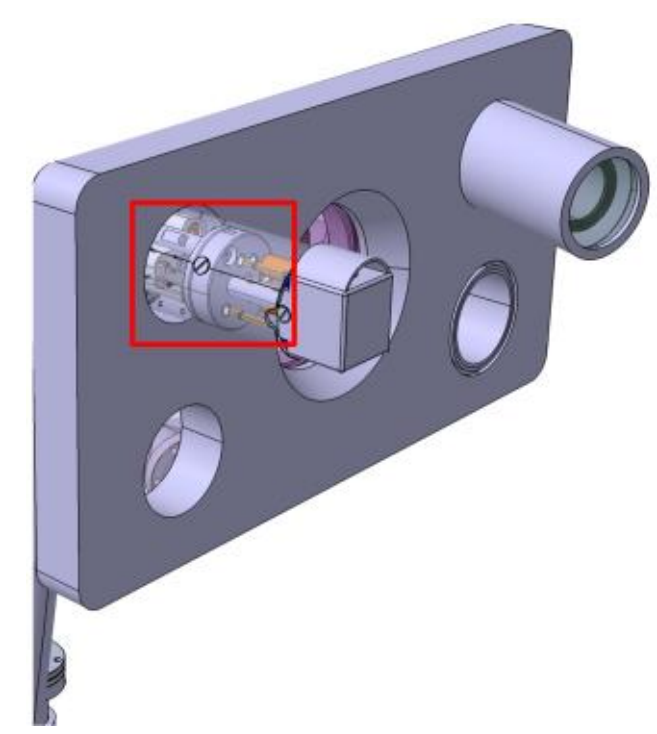

Figure 3. CAD image of the mid-plane gauge with the circular tube covering the leads of the gauge head (marked by red box).

Before operation starts, the communication between the filament power supply, NI controller and the FPGA must first be established, and remains active during the entire experimental session. The APG measurement is triggered seven seconds before the plasma discharge, to leave sufficient time to establish the requested value of $I^-$ before gas is let into the TCV vessel. After each TCV discharge, measured electron and ion currents are saved in MDS nodes of the TCV database. The neutral density and the corresponding neutral pressure are calculated according to Equation (1), using sensitivities that are determined in calibrations, section III.

## III. Calibration of the APG diagnostic

The sensitivity, Equation (1), is not a priori known and must be determined in a calibration. The sensitivity depends on the filament geometry, implicitly included in the surviving probability $S$ in Equation (2) [7]. As a result, recalibration of the gauges is for example necessary if the filament is deformed by the J×B force.

The APG sensitivities are determined in a cross-calibration with BPGs, using the duration of the gas flow to extend the BPG pressure range to lower values. The various steps of an APG calibration in TCV are described as follows.

### A. Design of the calibration pulses

The gauge sensitivity is determined from measurements of $I^+$ and $I^-$, using Equation (1), for a wide range of neutral pressures in dedicated calibration pulses in TCV.

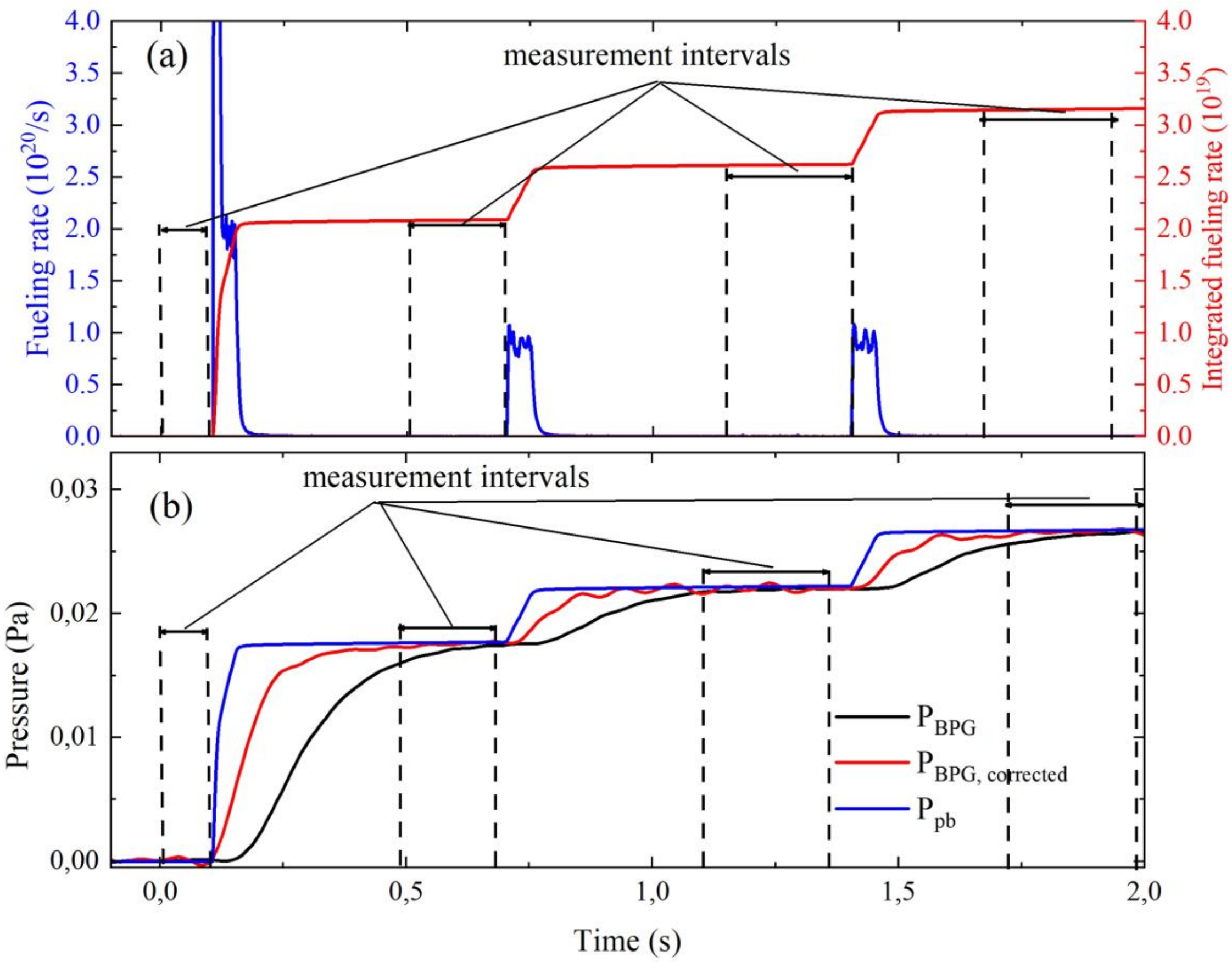

Figure 4. (a) Deuterium fueling rate (left axis) and its time integral (right axis) of TCV calibration pulse #77961. (b) Measured BPG pressure, $P_{BPG}$, corrected BPG pressure, $P_{BPG,corrected}$ (see section IIIB), and pressure based on particle balance. The four measurement intervals are marked. The magnetic field reaches 1.43 T at t= -0.05 s and remains constant until 2 s.

The APG calibration pulses feature the typical toroidal field of 1.43 T of a TCV plasma discharge, but no poloidal fields nor transformer currents. The gate valves of TCV's four turbo-molecular pumps are closed before deuterium gas ($D_2$) is injected into the vessel via a piezo valve. The neutral pressure distribution in the TCV vessel is expected to quickly equilibrate [5]. The $D_2$ gas is typically injected in three short pulses, separated by 0.6 s, leading to three discrete increases of the pressure and four measurement intervals with constant pressure, Figure 4(a). The first pressure measurement interval starts when the magnetic field reaches 1.43 T, typically at t=-0.05 s, but before the first gas pulse. The subsequent measurement intervals start 0.4 s after each $D_2$ pulse yielding interval durations of typically 0.2 s, where the measured quantities including currents and pressures are approximately constant and averaged, Figure 4(b). The delay between gas pulses and measurement intervals is determined by the time response of the BPGs. The APG calibration in TCV typically covers a pressure range from 0.5 mPa to over 200 mPa.

### B. Pressure estimation based on particle balance

Since APGs seek to extend the measurement range below the BPG limit, a direct cross-calibration with the BPGs is not sufficient. This difficulty is resolved by estimating the neutral pressure from the gas valve fluxes, based on the neutral particle balance.

The particle-balance estimate of the neutral pressure $P_{pb}$ is obtained from the integrated deuterium fueling rate:

$$P_{pb} = \frac{k_B T_g}{V_{TCV}} \int \Gamma_{D_2} dt, \tag{4}$$

where $k_B$ is the Boltzmann constant, $T_g$ is the gas temperature (~300 K), $V_{TCV}$ is TCV vessel volume and $\Gamma_{D_2}$ is the deuterium fueling rate. The value of $V_{TCV}$ is determined by comparing $P_{pb}$ with BPG pressures within the BPG measurement range.

The time delay and the slow time response of the BPG measurements can be corrected by deconvolving the system response with the following transfer function:

$$h(t) = \frac{1}{\tau_{resp}} \exp\left(-\frac{t-\tau_{delay}}{\tau_{resp}}\right) u(t-\tau_{delay}), \tag{5}$$

and the following relation for signal convolution:

$$P_{BPG}(t) = P_{BPG,corrected}(t) * h(t), \tag{6}$$

---

[5] The injected gas expands from the gas valve to the TCV chamber. The characteristic time of gas expansion in TCV chamber is estimated from $t_{expand} \sim \frac{\pi R}{v_{D2,th}}$, which is approximately 2 ms. Here $R$ is the major radius and $v_{D2,th}$ the thermal velocity of deuterium molecules at room temperature.

where * stands for a convolution, $h(t)$ is the Heaviside step function, and $P_{BPG,corrected}$ is the actual pressure at the sensitive surface of the BPG. The deconvolution algorithm computes $P_{BPG,corrected}$ from the measurements $P_{BPG}$ using values of $\tau_{resp}$ and $\tau_{delay}$ of 0.10 s and 0.04 s, respectively, that were determined in dedicated characterization experiments in TCV (not shown). The correction can effectively remove the effects of finite BPG time response and time delay on the measurement, Figure 4(b). However, the correction also amplifies the high frequency components of noise and, therefore, requires adequate low-pass filtering of the signal.

The comparison of the corrected BPG pressure with the particle balance estimate of pressure, Equation (4), in each measurement interval, yields an estimate of $V_{TCV}$ = 4.86 $m^3$.

### C. Treatment of the ion current offset

The determination of the APG sensitivity requires an accurate ion current signal. The ion current is expected to be approximately proportional to the neutral pressure, Equation (1), with potential offsets introducing systematic errors, in particular at low pressures. A separation of measured offsets into contributions from the residual pressure in the pumped vessel and intrinsic offsets reduces this systematic error.

The ion current offset has a non-negligible influence on the measurement at low neutral pressure (below 1 mPa) where the offset is comparable to the ion current signal. The evolution of the ion current before the first gas pulse has three phases, Figure 5. The ion current signal is non-zero even before the electron current emerges at approximately t= -3 s (phase 1). From t= -3 s to -1.2 s, the ion current slightly increases and saturates together with the electron current at a higher value (phase 2). The ion current further increases when the magnetic field is applied (phase 3). The phase 1 offset is found to be independent of the ion current gain. The phase 2 offset depends on the electron current, and phase 3 depends on both electron current and the magnetic field.

The phase 1 offset exists even without electron current, hence without electrons that may ionize any neutrals and is, therefore, an offset unrelated to the neutral pressure. The increments of the signals in phase 2 and 3 are likely caused by a residual pressure before the gas injection. Note, that the BPG signal is set to zero just before gas is injected into the torus. Therefore, any residual pressure is excluded in the BPG measurement. In phase 2, the electron current can already lead to the ionization of any residual gas and, hence, increase the ion current signal even if the electrons are not confined by a magnetic field. The phase 3 signal exceeds the phase 2 signal (both corrected for the phase 1 offset), by approximately one order of magnitude. This increase is presumably caused by the increase of the electron confinement in the direction perpendicular to the electric field by the magnetic field. In Equation (2), this increase of confinement is reflected by an increase of the electron oscillation number $f$. The increase of ion current with magnetic field is supported by tests of an APG in a vacuum chamber, where the effect of magnetic field saturates when field is higher than approximately 1.0 T, consistent with previous observations [7]. The phase 1 ion current offset is, therefore, subtracted when analyzing the calibration pulses and $I^+$ will henceforward refer to the offset corrected ion current measurement.

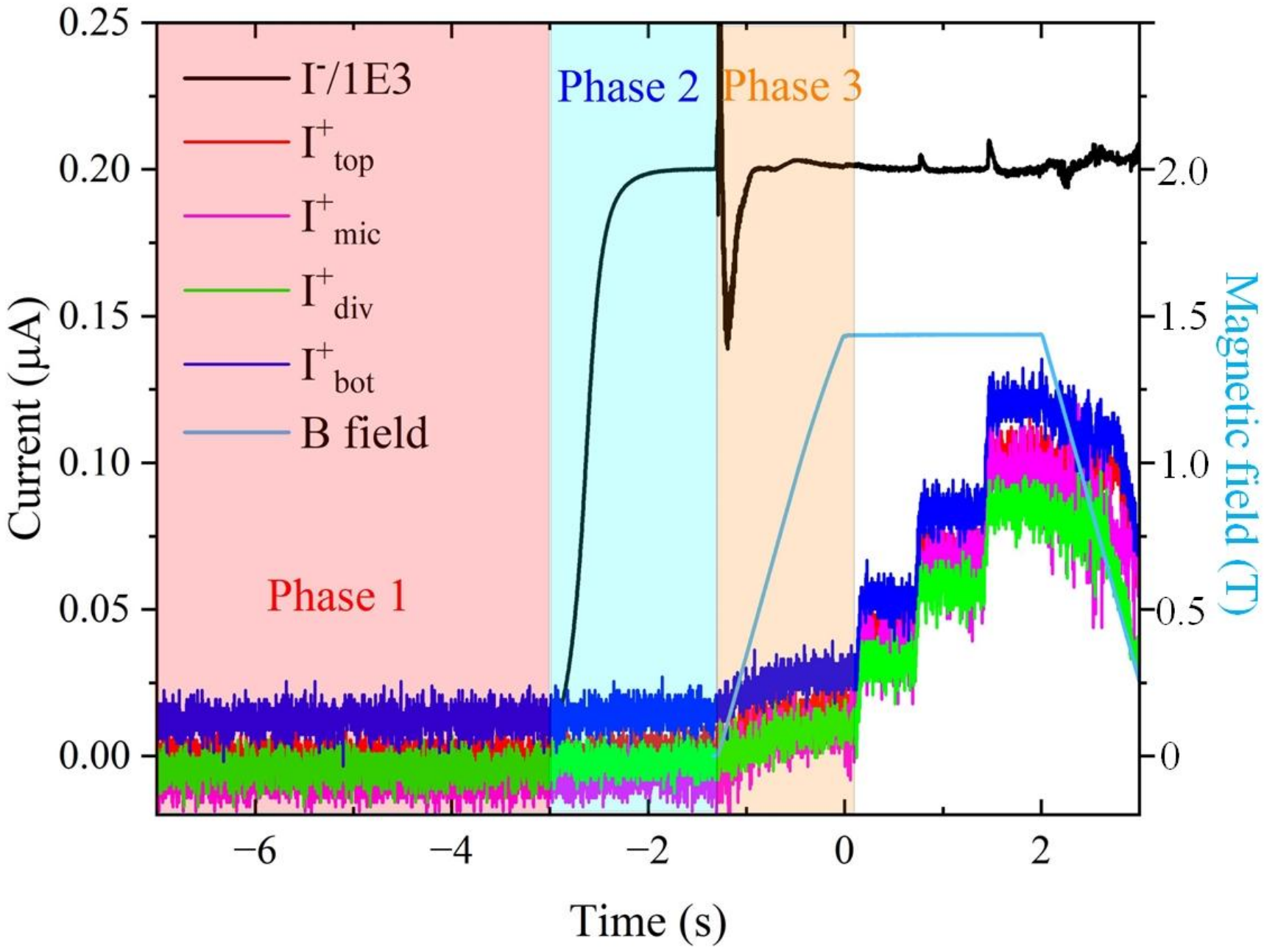

Figure 5. Normalized electron current of the APG 'mic', ion currents of four gauges, and the magnetic field of a low-pressure calibration pulse #77960, to better visualize the evolution of current offsets. The electron current is maintained at 200 μA during the measurement and the magnetic field ramps up to 1.43 T before descending. Ion current offsets in phase 1-3 are marked in shaded areas.

### D. Residual pressure correction

The phase 2 and 3 ion current signals discussed in section IIIC suggest a significant residual pressure, $P_{residual}$, before the deuterium injection in the calibration pulses, and should be added to the gas-balance pressure:

$$\frac{I^+}{I^- - I^+} = \frac{d}{k_B T_g}(P_{pb} + P_{residual}). \quad (7)$$

The value of $P_{residual}$ is obtained for each calibration pulse by fitting the measurement of $I^+$ and $P_{pb}$ of the four measurement intervals to Equation (7), Figure 6. The values of $P_{residual}$ of the four gauges obtained in a calibration pulse are similar, but vary among calibration pulses. The residual pressure is generally below 0.5 mPa, and is likely related to the duration of the time interval between closing the gate valves that separate the turbo pumps from the TCV vessel and the calibration measurements. The residual pressures of four APGs are averaged for a calibration pulse, and the total pressure $P_{tot} = P_{pb} + \overline{P}_{residual}$ of each measurement interval will be used for the sensitivity calculation in section IIIE.

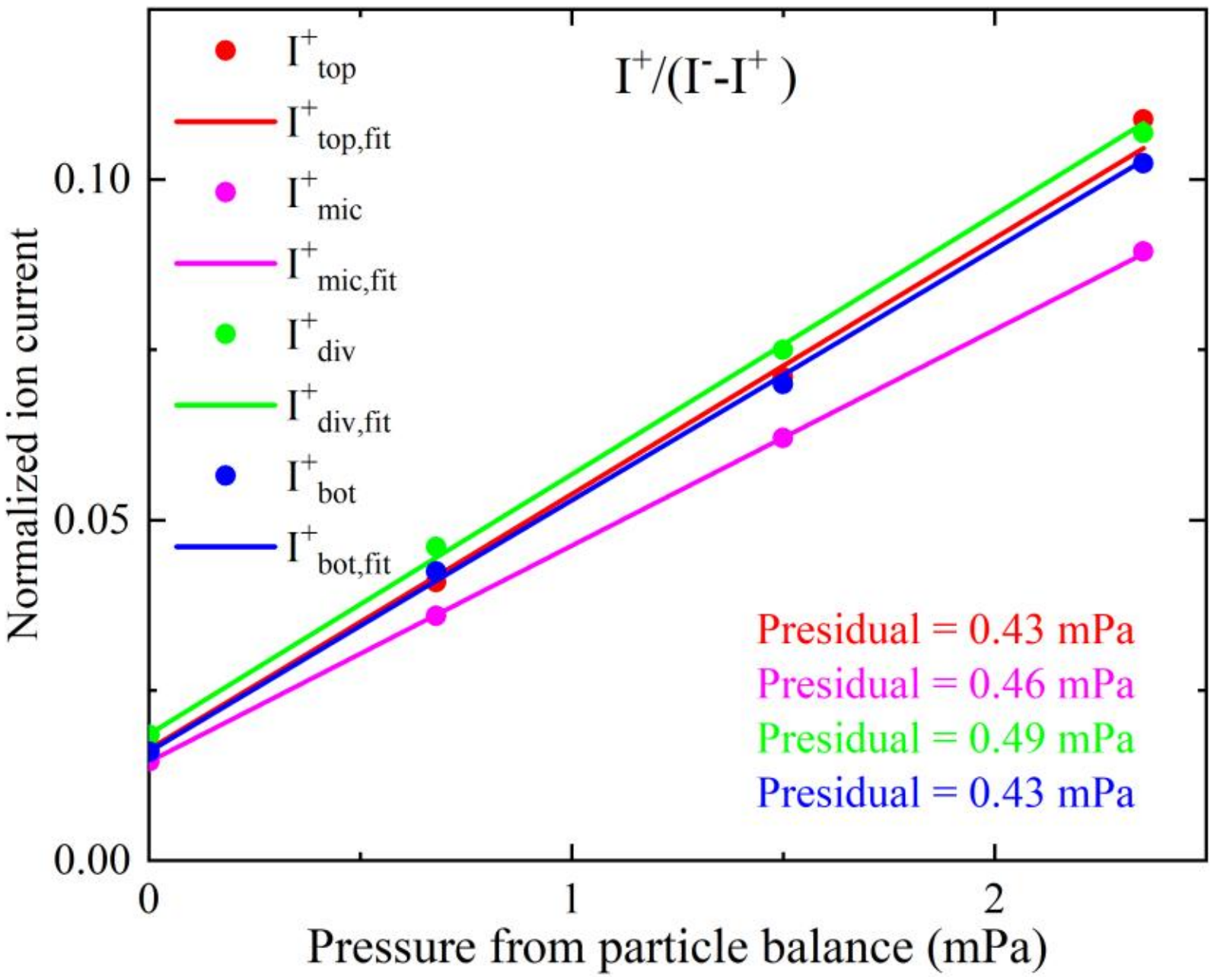

Figure 6. Averaged $\frac{I^+}{I^- - I^+}$ and $P_{pb}$ of the four measurement intervals of the calibration pulse #77960 for four APGs, and linear fits to determine the residual pressure based on Equation (7).

### E. Calculation of the sensitivity at the reference electron current

The sensitivities of each measurement interval in all calibration pulses are obtained using Equation (7). The residual pressure correction leads to a sensitivity that varies little with pressure above 10 mPa, but decreases below 10 mPa by up to 30%, Figure 7.

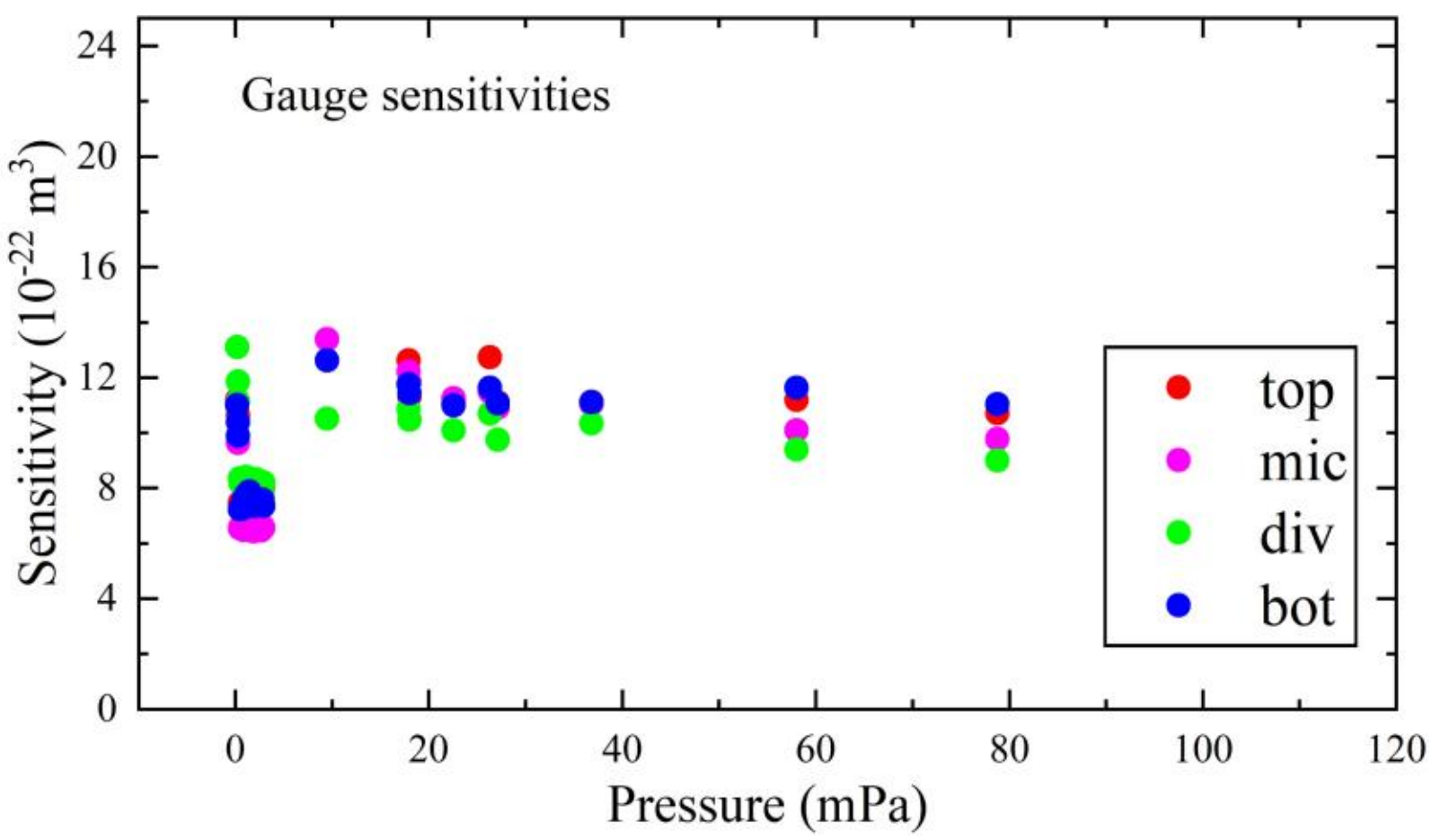

Figure 7. Dependence of gauge sensitivity on neutral pressure with the residual pressure correction, of the calibration session in August, 2023.

### F. Sensitivity with varying electron current

The calibration procedure introduced in section IIIE determines the gauge sensitivity at an electron current of 200 μA, which is typically used in TCV discharges. However, when plasma and neutral conditions change quickly, the value of $I^-$ may significantly deviate from the reference value before the controller can correct the filament heating and, hence the electron current. To improve the measurement precision during such $I^-$ fluctuations, the reference electron current is varied, and the obtained sensitivities, Figure 8, are parameterized with two parameters $d_0$ and $A$ [7]:

$$d = d_0(\frac{A}{I^-} + 1) \quad . \tag{8}$$

The pressure measured by an APG is then obtained from:

$$P_{APG} = \frac{I^+}{I^- - I^+} \frac{k_B T_g}{d_0(\frac{A}{I^-}+1)} \quad . \tag{9}$$

Since the available measurements are predominantly obtained at pressures above 10 mPa, the fit of the measured sensitivities to Equation (8) results in parameters $d_0$ and $A$ that overestimate $d$ and, hence, underestimate pressures below 10 mPa by up to 30% (see Section IIIE).

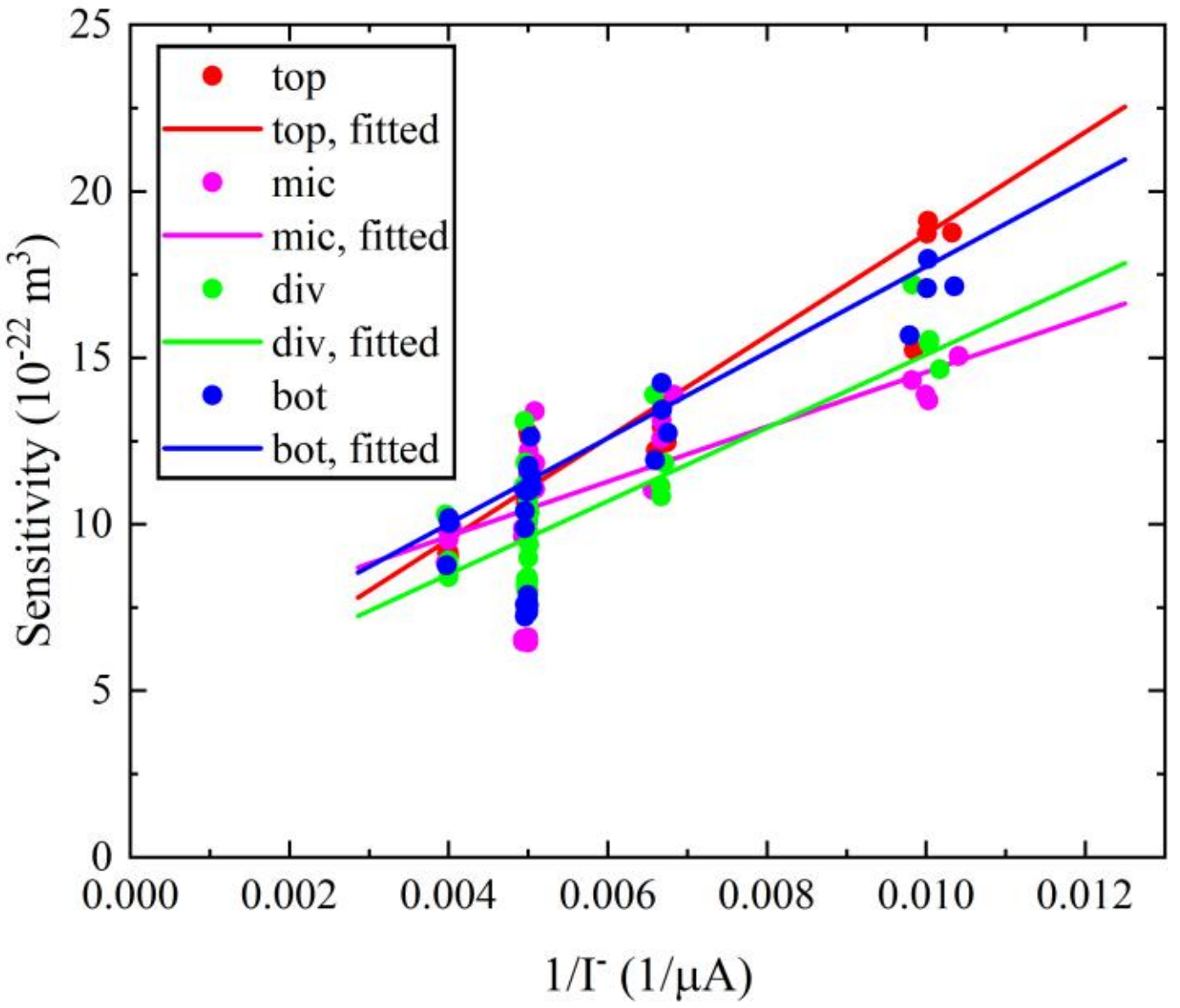

Figure 8. Dependence of the gauge sensitivity on the reciprocal of the electron current.

## IV. Evaluation of the APG performance

The performance of the APGs in TCV is evaluated by comparing them with the BPGs in the calibration pulses without plasma and during TCV plasma discharges. The advantages of APGs are described and causes of apparent discrepancies are analyzed.

### A. Comparison of APG and BPG measurements in TCV calibration pulses

The APG-measured pressure is compared with the BPG-measured pressure for two representative APG calibration pulses at low- and high-pressure levels, Figure 9. The APG measurement reveals a clear residual pressure, and responds quickly to $D_2$ injection with negligible time delay and a fast time response, Figure 9(a). The APG measurement in the low-pressure pulse confirms that the APG can resolve pressures as low as 0.5 mPa. The BPG pressure, both original and the corrected, can hardly resolve the first two pressure measurement intervals. The BPG pressure also exhibits strong irregular signal oscillations, amplified by the deconvolution procedure, which are observed at low-pressure levels.

The APG-pressure rises with the gas injection with a time delay, $\tau_{delay}$ (here defined as the time interval between the change of actual pressure and the change of the measured pressure), of approximately 3 ms, which is more than one order of magnitude lower than the delay of the BPG. The APG time delay likely due to the combination of molecule time of flight (approximately 2 ms) and averaging over chopping intervals (0.2 ms). The APG also features a response time (the speed for the measured pressure to approach and reach the actual pressure, here estimated from an exponential fit in a step response of the actual pressure) of less than 4 ms, which is more than an order of magnitude lower than the BPG response time. Therefore, APGs feature a much faster time response than BPGs, which is also shown in the high-pressure calibration pulse, where the APG pressures reach the constant level in a measurement interval after a $D_2$ pulse faster than the BPG corrected for $\tau_{delay}$, Figure 9(b).

The APG signal features a signal oscillation of less than 0.5 mPa, Figure 9(a), which does not vary significantly with the pressure, Figure 9(b). The APG pressure, however, exhibits "pressure jumps" in the third and fourth pressure measurement intervals of the high-pressure calibration pulse, where the measured pressure repetitively switches between multiple fixed levels during a measurement interval, suggesting transition between different modes. These jumps are caused by jumps in the ion current, a phenomenon that has previously been observed during APG operations on W7-X and AUG [18, 23]. Recent studies proposed that the ion current jump is likely related to the two-stream instability due to electron oscillations, a counter-streaming motion, between the filament and the ion collector [18]. Dedicated simulations of the APG plasma are needed to validate this hypothesis and seek potential mitigation approaches of the ion current jump. Though the ion current jump may degrade the accuracy of the diagnostic, it is currently not corrected due to the limited understanding of the underlying mechanism.

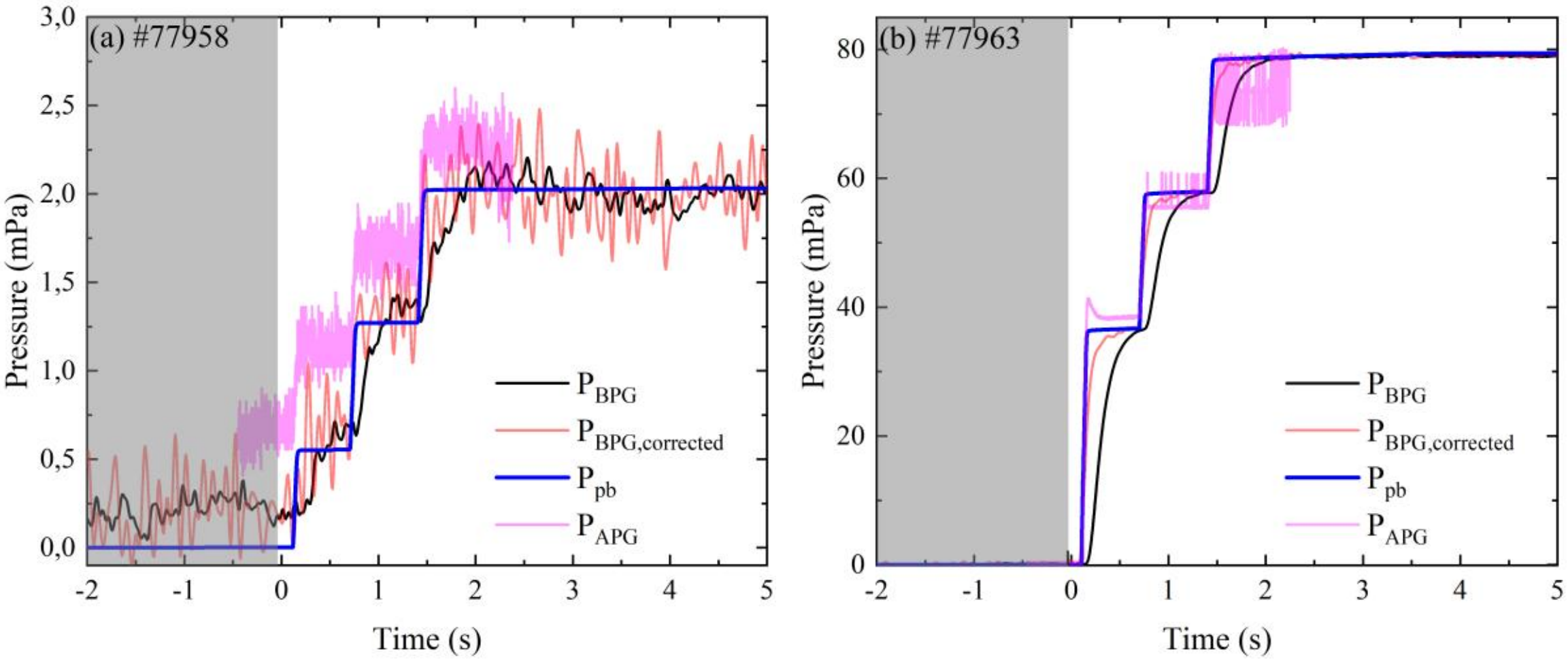

Figure 9. Measured BPG pressure, corrected BPG pressure, pressure based on particle balance, and the APG pressure of the TCV calibration pulse #77958 and #77963. Only outer mid-plane gauges are compared. The time when the magnetic field reaches maximum is marked by the shaded area.

### B. Comparison of APG and BPG measurements in TCV plasma discharges

To evaluate the APG performance in TCV plasma discharges, the pressure measurements in a series of Ohmic density ramp discharges by three APGs ('div', 'bot', and 'mic') are compared with the pressure measured by the BPGs installed in the same ports, Figure 10. The selected discharges have a plasma current of 250 kA and were performed in TCV's short inner, long outer (SI-LO) baffle configuration.

Prior to a TCV plasma discharges, the vessel is filled with deuterium gas starting at t = -0.1 s. The gas break down is induced through the application of a large loop-voltage at approximately t = 0 s. After the divertor configuration is fully developed at t = 0.5 s, the plasma density is increased via gas fueling until the discharge disrupts at t = 1.1 s. During the entire discharge, the APG measurements feature a faster time response and a lower time delay than the BPGs, Figure 10, as in the non-plasma pulses, Figure 9. While APG and BPG measurements generally agree before and after the plasma discharge, i.e. when the vessel is primarily filled with deuterium gas, the APG-measured pressures are systematically higher in the presence of plasma in the vessel. The difference is clearly visible on the divertor gauges, Figure 10(a), (b), whereas in the outer mid-plane, the BPG cannot resolve the pressure very well as the neutral pressure during plasma discharge is below 5 mPa, Figure 10(c). The BPG measurement exhibits strong oscillations while the APG can resolve the mid-plane pressure below 5 mPa. During the density ramp, the corrected BPG pressures in the divertor CFR and PFR are higher than the uncorrected BPG measurements, due to the slower time response and considerable time delay of BPGs [24]. However, this correction alone cannot fully explain the discrepancy between APG and BPG measurements.

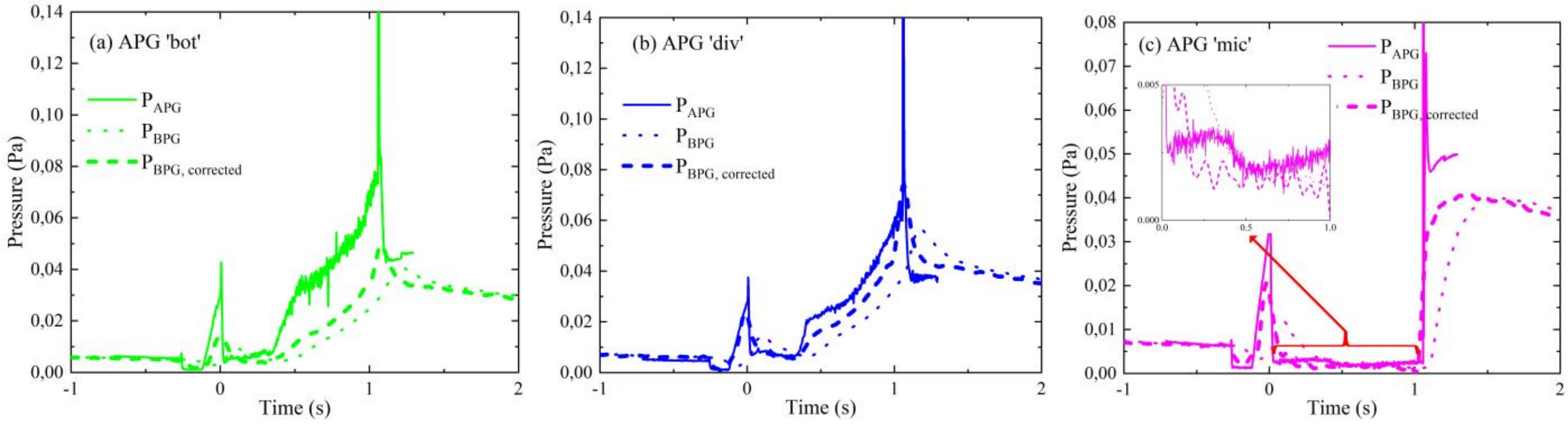

Figure 10. Example of the APG, measured and corrected BPG pressures for a TCV discharge with a density ramp (#78188). The APGs and BPGs are located at the same ports in the (a) divertor common flux region, 'bot', (b) private flux region, 'div', and (c) outer mid-plane, 'mic'. The insert in (c) highlights the measurements between t = 0 s and 1 s.

A comparison of APG and BPG measurements in several TCV discharges with different density ramp rates shows the extent of the disagreement, Figure 11. During these discharges, APGs systematically measure a pressure that is approximately 120% and 33% higher than the BPG measurement, for divertor CFR and PFR gauges, respectively. The pressure at the outer mid-plane is low and is beyond the BPG measurement limit, hence the mid-plane BPG pressure is scattered and no clear pressure dependence on the APG pressure is found.

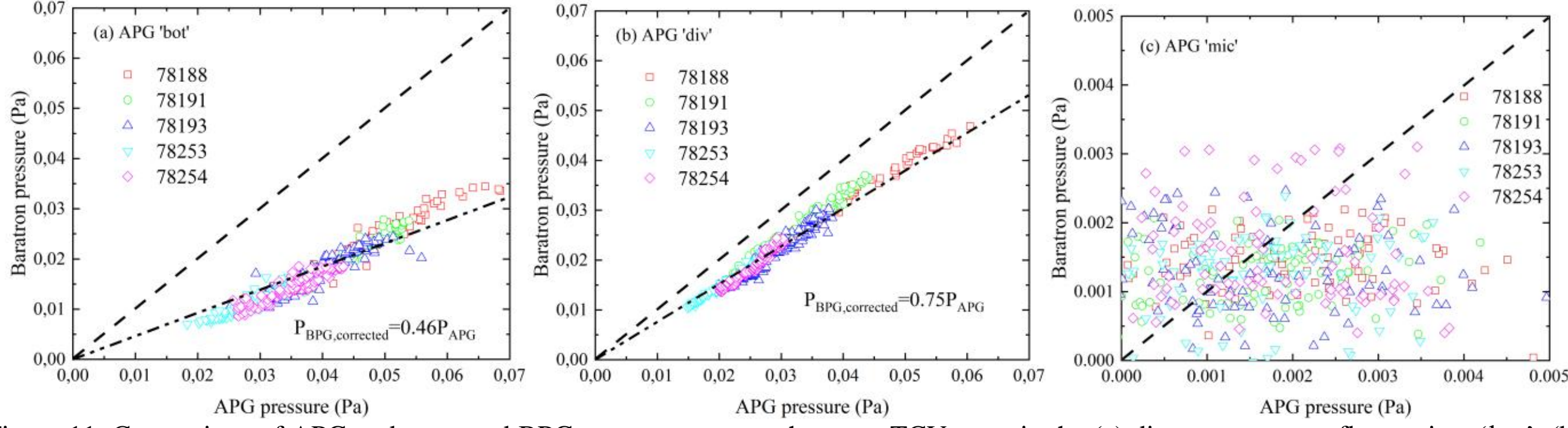

Figure 11. Comparison of APG and corrected BPG measurements at the same TCV ports in the (a) divertor common flux region, 'bot', (b) divertor private flux region, 'div', and (c) outer mid-plane, 'mic'. Five discharges with density ramp rates increasing from 3.5 $m^{-3}s^{-1}$ to 7.8 $m^{-3}s^{-1}$ are selected for the comparison.

### C. Analysis of the pressure measurement discrepancies

The remaining discrepancies between APG and BPG measurements can have several reasons. A key observation is that the discrepancies in the measured pressures only occur in the presence of plasma, when neutrals in the vicinity of the gauge orifice are only partially recombined to molecules and atoms may have a non-Maxwellian distribution. One possible explanation is that the neutral fluxes into gauges located in the same port are not uniform during a plasma discharge.

To validate this hypothesis, SOLPS ITER 3.0.8 code package is used to model the deuterium neutral distribution during a plasma discharge in the TCV gauge ports. The SOLPS-ITER code couples the 2D fluid transport code B2.5 and the 3D kinetic neutral transport code EIRENE [25]. The code employs axisymmetric geometry, including axisymmetric gauge port structures, and can, therefore, only provide a qualitative assessment. The employed magnetic equilibrium corresponds to the configurations used in the density ramp experiments described in Section IVB. Considered plasma species are deuterium and carbon, and the deuterium is fueled from the outer divertor common flux region. The simulation setup such as the boundary conditions, considered reactions, and transport coefficients, corresponds to previous work using the same magnetic equilibrium [26, 27]. The adopted magnetic equilibrium, plasma and neutral grid, and gauge locations are shown in Figure 12. The extended gauge port structures at the top, the high-field side floor, the low-field side floor, and the outer mid-plane of the TCV chamber are based on realistic TCV port geometry but are axisymmetric due to code limitations.

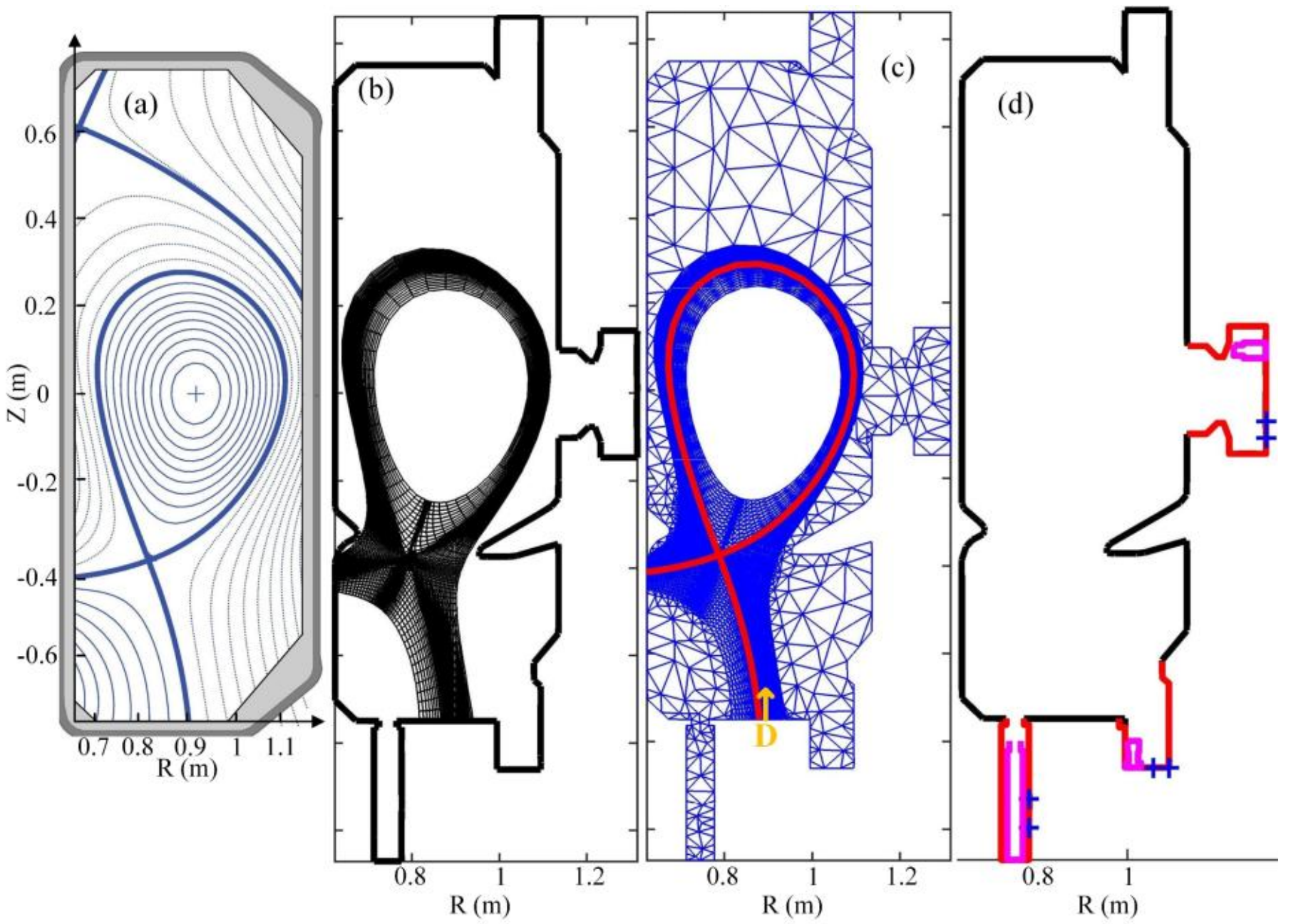

Figure 12. (a) Employed TCV magnetic equilibrium. (b) B2.5 grid (c) EIRENE grid, where the deuterium fueling location is marked by an arrow. (d) Locations of the APG probe head (magenta) and the openings for the BPG extension tubes (blue).

Synthetic pressure measurements are obtained from SOLPS-ITER simulations based on the assumption that the neutral distribution inside the gauge volume is dominated by thermalized deuterium molecules, Equation (3):

$$P_{syn,gauge} = \sqrt{\pi m_D k_B T_{gauge}}(\Gamma_D^{in} + 2\Gamma_{D2}^{in}) \ , \tag{10}$$

where $m_D$ is the deuterium atomic mass, and $T_{gauge}$ the gauge temperature, typically assumed to be 300 K. It has, hitherto, been assumed that directed neutral fluxes, $\Gamma_D^{in}$ and $\Gamma_{D2}^{in}$, within TCV ports are uniform and correspond to the neutral flux into the port, an assumption that was supported by previous SOLPS-ITER simulations of the BPG extension tube [28]. The assumption is challenged by basing the synthetic pressure measurements, equation (10), on the simulated neutral fluxes onto the APG probe head and the opening for the BPG extension tube, Figure 12.

The neutral pressures from synthetic APG measurement exceeds the synthetic BPG measurement in the divertor and outer mid-plane by 45%, 40%, 66%, respectively, Table 1.

Table 1. Synthetic pressure measurement from SOLPS, Equation (10)

| Pressure (mPa) | div | bot | mic |
|---|---|---|---|
| APG | 15.0 | 10.3 | 0.5 |
| BPG | 10.3 | 7.4 | 0.3 |

The synthetic pressure values suggest that the neutral fluxes towards the gauges within the same port are not uniform. SOLPS-ITER simulations predict that neutral density in the gauge port is lower near the port entrance, and increases away from the port entrance, particularly in corner regions without direct line of sight facing the SOL plasma, Figure 13. Note that the neutral density and energy both affect the neutral flux into the gauge, so that the relative importance of values in Table 1 is not always reflected in Figure 13.

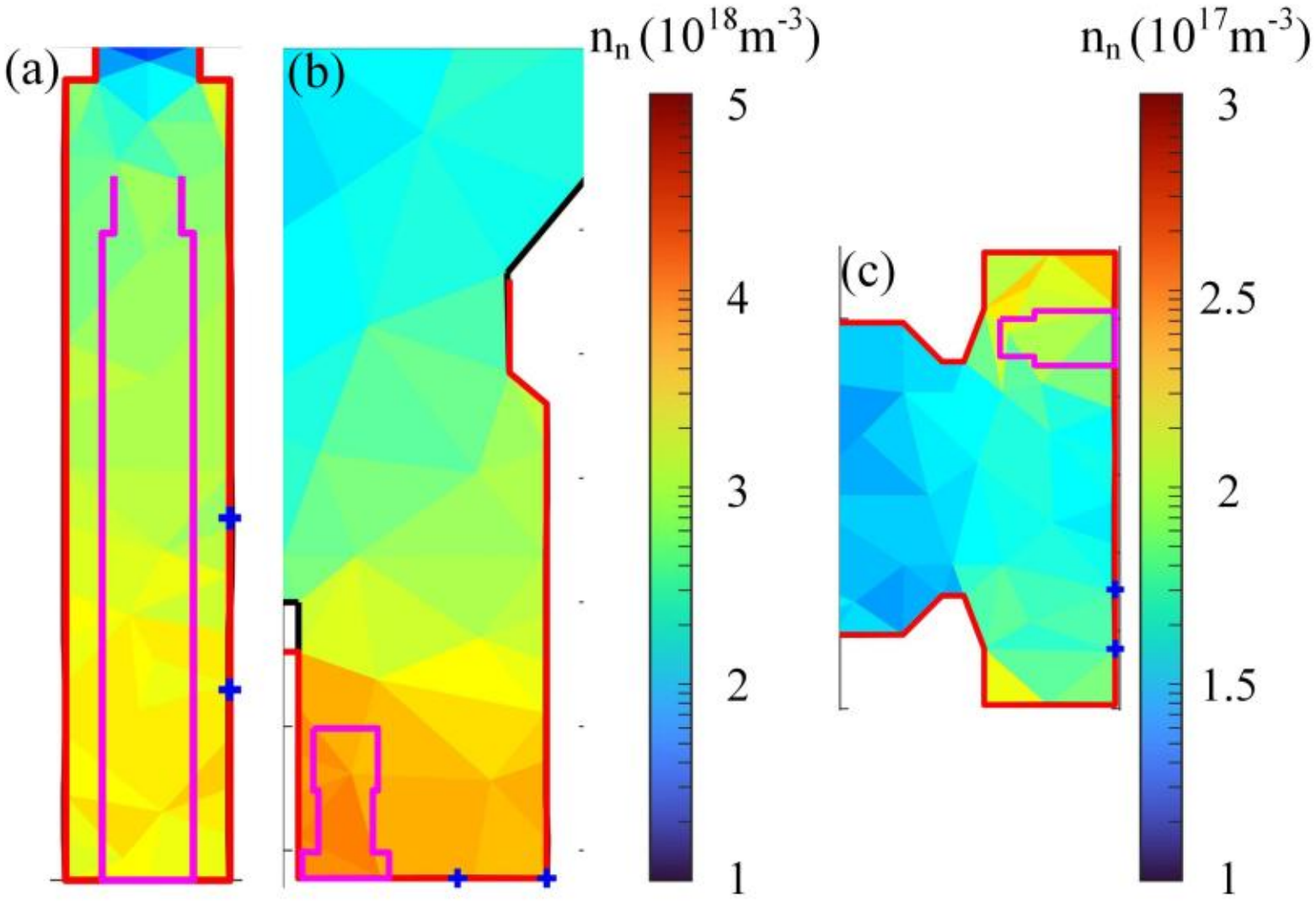

Figure 13. Total neutral deuterium density ($n_{D,tot} = n_D + 2n_{D2}$) distribution in the three gauge ports predicted by SOLPS-ITER simulation. (a) APG 'div'. (b) APG 'bot'. (c) APG 'mic'. The magenta structures are transport to the neutrals and are added only to evaluate the neutral fluxes in the simulated volume. The blue crosses mark the entrance to the BPG extension tubes.

The SOLPS-ITER simulation reproduce the direction and the order of magnitude of the difference between particle fluxes into the APGs and BPGs needed to explain the apparent discrepancies. A quantitative comparison would, however, require a realistic 3D port geometry and a precise description of the divertor plasma in the simulation.

In conclusion, the APGs installed in TCV exhibit significantly faster time response and lower time delay than the TCV BPGs. There are discrepancies between the BPG and APG measurement, due to two reasons: (1) BPGs have slower time response and larger time delay than the APG. (2) Neutral pressure distribution is nonuniform in the gauge port during plasma discharge. Further investigations are expected in future APG operations in TCV to better interpret the two types of pressure measurement.

## V. Conclusion

ASDEX-type pressure gauges have recently been installed in the TCV tokamak to probe the neutral distribution at the top, the high-field side floor, the low-field side floor, and the outer mid-plane of the TCV vessel. The APGs were integrated into the TCV discharge cycle. The APGs are calibrated for $D_2$ using calibration pulses. The pressure is estimated from the gas injection rate and the neutral particle balance based on a cross calibration with baratron gauges at sufficiently high pressures. Measured ion current must be corrected for an offset unrelated to neutral pressure, and the estimated residual pressure that is present before the programmed gas flow. Sensitivities are determined at different pressure and electron current levels. The calibration procedure yields sensitivities which do not greatly vary with the neutral pressure and vary linearly with the reciprocal of electron current. The performance of the APGs and the BPGs located at the same TCV ports are compared, showing a faster time response, lower time delay, and an extension of the measurement range to pressures below 5 mPa for the APGs. The discrepancy between APG and BPG measurement can be explained by differences in the time delay and time response, and a nonuniform neutral distribution in the gauge ports during the TCV discharge. The initial APG operations in TCV are promising, and validate that APGs should be reliable for pressure measurement in the next TCV upgrade.

## Acknowledgments

This work has been carried out within the framework of the EUROfusion Consortium, via the Euratom Research and Training Programme (Grant Agreement No. 101052200— EUROfusion) and funded by the Swiss State Secretariat for Education, Research and Innovation (SERI). Views and opinions expressed are however those of the author(s) only and do not necessarily reflect those of the European Union, the European Commission, or SERI. Neither the European Union nor the European Commission nor SERI can be held responsible for them. This work was supported in part by the Swiss National Science Foundation. The authors would like to thank Dr. Felix Mackel, Dr. Michael Griener from Max–Planck Institute for Plasma Physics, and Mr. Wolfgang Noack from Arbeitsgruppe Weltraumphysik und-technologie for helpful discussions and continued support.